\documentclass[pdflatex,sn-mathphys-num]{sn-jnl}

\usepackage{graphicx}%
\usepackage{multirow}%
\usepackage{amsmath,amssymb,amsfonts}%
\usepackage{amsthm}%
\usepackage{mathrsfs}%
\usepackage[title]{appendix}%
\usepackage{xcolor}%
\usepackage{textcomp}%
\usepackage{manyfoot}%
\usepackage{booktabs}%
\usepackage{algorithm}%
\usepackage{algorithmicx}%
\usepackage{algpseudocode}%
\usepackage{listings}%
\usepackage{xspace}
\usepackage{threeparttable}
\usepackage{caption}
\usepackage{subcaption}
\usepackage{tikz}
\usetikzlibrary{positioning, arrows.meta, fit, backgrounds, calc}

\let\orcidlogo\undefined 
\usepackage{orcidlink}

\newcommand{\finding}[2]{%
  \par\medskip\noindent
  \fbox{\parbox{\dimexpr\linewidth-2\fboxsep-2\fboxrule\relax}{%
    \small\textbf{#1}\par\smallskip #2}}%
  \par\medskip}

\newcommand{\researchq}[1]{\textbf{RQ#1}\xspace}

\theoremstyle{thmstyleone}%
\theoremstyle{thmstyletwo}%

\theoremstyle{thmstylethree}%

\begin{document}

\title[Article Title]{Code-Grounded Agentic Explanations for Static Analysis Vulnerability Findings}


\author*[1]{\fnm{Oshando} \sur{Johnson}\,\orcidlink{0009-0001-1884-7969}}\email{oshando.johnson@iem.fraunhofer.de}

\author[1,2]{\fnm{Ranjith} \sur{Krishnamurthy}\,\orcidlink{0000-0002-0906-5463}}\email{ranjithk@hni.uni-paderborn.de}

\author[1,3]{\fnm{Rohith Kumar} \sur{Shanmuganathan}\,\orcidlink{0009-0006-1654-0660}}\email{rohith.shanmuganathan@uol.de}

\author[4]{\fnm{Goran} \sur{Piskachev}\,\orcidlink{0000-0003-4424-5838}}\email{gpiskach@amazon.com}

\author[1,2]{\fnm{Eric} \sur{Bodden}\,\orcidlink{0000-0003-3470-3647}}\email{eric.bodden@uni-paderborn.de}

\affil*[1]{\orgname{Fraunhofer IEM}, \orgaddress{\city{Paderborn},  \country{Germany}}}

\affil[2]{\orgname{Heinz Nixdorf Institute, Paderborn University}, \orgaddress{\city{Paderborn}, \country{Germany}}}

\affil[3]{\orgname{Carl Von Ossietzky Universität Oldenburg}, \orgaddress{\city{Oldenburg},  \country{Germany}}}

\affil[4]{\orgname{Amazon Web Services}, \orgaddress{\city{New York},  \country{United States}}}


\abstract{Static application security testing (SAST) tools surface vulnerabilities early and their verdicts are rule-based and auditable. Yet developers routinely dismiss findings they cannot interpret. Large language models can supply the missing explanation, but designs that also hand the model the detection give up the auditability that makes static analysis defensible. We keep detection with the analyser and give the model only the explaining. A preliminary user study with 20 participants and a workshop with 19 practitioners yielded five requirements. Guided by these we redesigned SAFE, an open-source IntelliJ IDEA plugin that explains findings from any analyser reporting in the Static Analysis Results Interchange Format. It tiers explanations by demand, ties them to the reported data flow, and replaces the single-prompt backend with an agent that must consult the repository and the rule that fired before answering. A within-subjects crossover experiment with 24 participants compared the redesign against its predecessor. Perceived usability rose from 77.3 to 85.0 on the System Usability Scale. Actionability gained most among the explanation-quality items, the aspect both earlier studies flagged as weakest, and the odds of a correct remediation were 2.6 times higher. Triage accuracy and time per finding were indistinguishable, which we read as uninformative rather than as equivalence. Trust was the one measure that did not move, and the closing interviews indicate that grounding made findings easier to understand rather than easier to believe. Participants also drew a boundary no requirement anticipated, wanting the tool to find and explain while they judge and apply.
}


\maketitle

\section{Introduction}\label{sec:intro}

Static application security testing (SAST) tools detect potential security vulnerabilities early in the development process, and their verdicts are rule-based, reproducible, and auditable. Yet SAST tools remain underused in practice~\cite{10.1145/3691621.3694947}, and a major reason is the poor usability and quality of the explanations they provide. 
A large-scale study of the usability criteria addressed by static analysis tools~\cite{10.1145/3533767.3534374, nachtigall2019explaining} found that many tools' terse rule identifiers and generic messages fail to meet basic usability criteria. Some of the usability criteria include warning messages that expose the tool's internal reasoning, explanations about the consequence of an issue, or the identification of the underlying weakness. Developers who cannot tell why a warning was raised, whether it is real, or how to fix it tend to dismiss it.

Large language models (LLMs) are capable of closing this explanation gap. However, most existing approaches rely on the LLM for both detecting and explaining vulnerabilities~\cite{llmvulexp, dlap_llm}. This gives up the auditability that makes SAST suitable for security-critical processes, and systematic comparisons report that models and analysers fail in different ways rather than one superseding the other~\cite{gnieciak2025large}.
The concern is not only that detection becomes unauditable. A model's stated reasoning is itself weak evidence of how it reached an answer, because chain-of-thought explanations can be steered by features of the input that the model never mentions while remaining fluent and plausible, with accuracy dropping by as much as 36\,\% under such manipulations~\cite{turpin2023language}. An explanation is therefore most trustworthy when the reader can check it against something outside the model. Commercial offerings already pair a language model with an analyser, though they aim at repair rather than explanation. GitHub's Copilot Autofix generates suggested patches for CodeQL alerts~\cite{copilot-autofix}, and Snyk's DeepCode AI Fix does the same for Snyk Code, using program analysis to narrow what the model attends to~\cite{deepcodeaifix}. Both are tied to their vendor's own detection engine, so neither travels to an analyser a team already runs, and in both the explanation is a by-product of proposing a fix rather than the artefact being produced. Our concern is the reverse. A developer must first decide whether a finding is real, which is a judgement a patch does not help with, and that judgement needs an account of the analyser's reasoning rather than a suggested diff.

We instead pursue a hybrid approach in which the SAST tool detects the vulnerability and the LLM explains it. Prior work has similarly applied models to analyser output, most closely by inspecting the warnings a SAST tool has already produced in order to filter false positives and describe them in natural language~\cite{auto_inspect_warnings_llm_how_far_are_we}. A second line of work changes what a model has access to rather than what it is asked. Agentic systems give the model a set of tools and let it decide which to call and in what order, so that it gathers evidence over several steps instead of answering from a single fixed prompt~\cite{yao2023reactsynergizingreasoningacting}. In software engineering this has been applied to writing and repairing code, where an agent navigates a repository, runs commands and edits files until a task is complete~\cite{3737916.3739517, 3692070.3694124, aider}. The same mechanism is useful for a different purpose. An explanation that has to refer to a developer's own code cannot be assembled from a snippet chosen in advance, because what is relevant depends on where the untrusted value came from and what the flagged rule checks, and neither is known until the model has looked. Retrieval decided by the model is therefore a better fit than a prompt fixed before it starts.

In previous work, we introduced the first version of the Static Analysis Findings Explainer (SAFE), an IntelliJ IDEA plugin that uses LLMs to explain the vulnerabilities detected by SAST tools, and evaluated the completeness of the generated explanations with software security trainers~\cite{safe-ex-ase}. Two limitations remained. The trainers judged whether an explanation read the way an expert would write one, which is not the same question as whether it helps a developer act, and they judged it in a demonstration rather than while triaging findings in their own project, so the study could not observe what a developer decided, how long it took, or whether the fix they proposed was right. The explanations were also assembled from the analyser's output and a snippet extracted around the finding. The model never saw the definition of the rule that fired and could not follow a data flow across files, which left it paraphrasing textbook knowledge of the weakness class.

In this paper we extend that work from evidence to redesign. A preliminary user study with 20 participants compared an initial prototype against a baseline SAST tool, establishing that developers clearly prefer LLM-generated explanations while exposing the limitations above. A structured workshop with practitioners from industry and research then supplied what such a tool must do in practice, and we consolidate both studies into five takeaways and five requirements. The redesign follows from those requirements. Explanations are tiered so that the essentials are visible at a glance, each one links to the data flow the analyser reported, and an agentic backend grounds it in the developer's own codebase by reading and searching the project before answering. Unlike a general-purpose coding agent, it is constrained to explain the analyser's existing verdict and is forbidden from detecting or patching anything itself. 

We evaluate the result against its predecessor in a within-subjects crossover experiment with 24 participants. The two builds differ in several respects at once, so the comparison establishes that the redesign performs differently rather than isolating which of its parts is responsible. Perceived usability improves, the widest perception gain is actionability, which both formative studies had identified as the weakest part of the earlier explanations, and participants propose better fixes. Triage accuracy is indistinguishable between the tools, and we read that null as uninformative rather than as evidence of equivalence. One measure stays flat, and it is the one the design was expected to move. Participants rate the evidence trail as no more worthy of trust than the text it replaced, even though it is the part of the redesign that can be checked. The closing interviews locate the reason. Participants anchor their trust to their own reading of the code, and they value the trace for taking them to it quickly. Those interviews also surface a boundary no requirement anticipated, since participants want the tool to find and explain while they judge and apply.

In summary, this paper makes the following contributions:

\begin{itemize}
    \item two formative studies of LLM-generated vulnerability explanations, a user study with 20 participants and a workshop with 19 practitioners, consolidated into five takeaways and five requirements;
    \item a requirements-driven redesign of SAFE whose architecturally novel component is an agentic backend constrained to explain a SAST tool's existing verdict;
    \item the open-source SAFE IntelliJ IDEA plugin~\cite{plugin}, which delivers these explanations directly in the Integrated Development Environment (IDE) for any SARIF-producing SAST tool;
    \item a within-subjects crossover experiment with 24 participants comparing the redesigned system against its predecessor on the same findings.
\end{itemize}

The remainder of this paper is organized as follows. Section~\ref{sec:elicitation} reports the preliminary study at the foundation of our work and the industry workshop we conducted to obtain additional feedback. Section~\ref{sec:safe-agent} describes the redesigned plugin and the agentic explanation approach. Section~\ref{sec:evalaution} reports the evaluation of the approach, and Section~\ref{sec:discussion} discusses the results. Section~\ref{sec:related-work} surveys related work, and Section~\ref{sec:conclusion} concludes the paper.

\section{Formative Studies and Requirements}
\label{sec:elicitation}

Before redesigning the plugin, we drew on two formative studies. A preliminary user study in Section~\ref{sec:pre-study} established how developers read LLM-generated explanations of SAST warnings, and an industry feedback workshop in Section~\ref{sec:industry-workshop} showed how such explanations have to behave in a professional setting. Section~\ref{sec:takeaways} consolidates both into five takeaways, from which Section~\ref{sec:requirements} derives the requirements for the redesign.

\subsection{Preliminary User Study}
\label{sec:pre-study}

Our work builds on a preliminary user study of an initial plugin prototype. The IntelliJ IDEA plugin listed the vulnerabilities reported by a SAST tool, and on request a controller assembled a prompt from the warning message, the affected code snippet, and surrounding project context, sent it in a single request to GPT-4, and rendered the explanation next to the finding. The study asked how large language models can be used to comprehensibly explain warnings generated by static analysis tools.

\subsubsection{Study Design} 
We conducted a within-subjects study~\cite{within_subject} with counterbalanced tool order~\cite{usability_uxDesign_evaluation}, comparing the prototype against Qodana~\cite{Qodana}, the JetBrains SAST plugin for IntelliJ IDEA, so that both conditions shared the same IDE surface. We chose a user study over automatic text-similarity metrics such as BLEU or ROUGE because those do not assess explanations from the perspective of the users they are meant to serve~\cite{whom_do_explanation_serve}, and over an automated judge, whose agreement with human raters is itself contested~\cite{chiang-lee-2023-large}. We hypothesized that its explanations help developers identify vulnerabilities more effectively (H1) and more efficiently (H2) than the baseline, and that developers rate them higher (H3). As material we scanned the OWASP Benchmark~\cite{owasp_benchmark_git} with Semgrep~\cite{Semgrep} and randomly selected two findings for each of five injection vulnerability types, four of which appear in MITRE's CWE Top 25~\cite{mitre_cwe_top_25}. In an online survey, 20 students and professionals judged from screenshots whether a reported vulnerability was real, once per tool for each type, and rated both tools on a 6-point scale. Because the ratings are ordinal and the sample is small, we report Wilcoxon signed-rank tests throughout.

\subsubsection{Results and Discussion} 
Effectiveness and efficiency remained inconclusive. Measured as the proportion of correctly classified warnings, the ten participants who started with the prototype reached a mean of 0.52 with it and 0.56 with Qodana, while the ten who started with Qodana reached only 0.30 with the prototype and 0.24 with Qodana. Within each group the difference between the two tools was not significant, with $p = .222$ for the group that started with the prototype and $p = .833$ for the group that started with Qodana. Effectiveness was therefore dominated by the order in which participants saw the tools rather than by the tools themselves, which left H1 unresolved. The survey setting, in which reading long explanations dominates task time, left H2 unsupported.

Satisfaction clearly favored the prototype and supports H3. Participants rated it higher on every feedback item, with a per-participant mean of 4.24 across all explanation-quality items versus 3.45 for Qodana and 13 of the 20 participants preferring it. On the matched clarity item, the prototype reached 4.45 versus 3.45 ($p = .022$, pooled across both order groups). It scored highest on the clarity of the explanation, on referring to the actual vulnerable code, and on improving understanding over the SAST tool (each $M = 4.45$), and lowest on the level of detail of the mitigation steps ($M = 3.85$).

The study also exposed limitations of the prototype. Participants found the explanations too long and asked for output that states the presence of a vulnerability directly and reveals cause and mitigation only on demand. The fixed explanation structure sometimes pushed the model to explain false positives as if they were real vulnerabilities, and individual respondents noted explanations that contradicted the SAST tool's verdict. Mitigation detail was the lowest-rated aspect, with participants asking for corrected code examples. The prompt was assembled from the warning message and code snippet, so the model was not left to reason about the finding entirely on its own. It never saw the definition of the rule that had fired, however, and it could not read beyond the extracted code snippet, so its account of why the analyser objected rested on the rule's message rather than on the conditions the rule actually checks, and vulnerabilities spanning several files could not be adequately explained. Finally, the screenshot-based survey could not capture how developers interact with explanations in a live IDE.

\subsection{Industry Feedback Workshop}
\label{sec:industry-workshop}

The insights of Section~\ref{sec:pre-study} guided the first version of the SAFE plugin, which we evaluated with software security trainers in a workshop paper~\cite{safe-ex-ase}. Both evaluations were indirect, since the preliminary study showed screenshots in an online survey and the trainer evaluation targeted an educational rather than an industrial setting. We therefore conducted a structured feedback session during the consortium meeting of a research project, the first setting in which the plugin was used by its actual target group, in order to understand the concerns of practitioners who work with security findings and to derive requirements for the second version.

\subsubsection{Workshop Setup and Participants}
The session followed a World Café format that combined moderated discussion, hands-on interaction, and a short survey. An opening discussion covered the pain points participants face when acting on vulnerabilities, the support they expect from an explanation tool, and the factors that would make such a tool unusable. Participants then used the plugin on real findings and assessed the generated explanations in context, so that feedback rested on actual interaction rather than on abstract expectations, and a closing discussion collected improvement ideas and conditions for practical use. 

The consortium meeting included 19 participants from industry and research, among them security experts (1), product team members  (3), developers (5), and researchers (9). Six of them completed a two-part survey with Likert items and open text fields, four in the role of developer, one as security expert, and one as researcher. Five rated their experience with secure programming as medium and one as rather extensive, and three found it rather difficult to understand the technical background and impact of vulnerabilities.

\subsubsection{Results and Discussion}
On a 5-point scale, four respondents rated the plugin as rather useful, one as medium, and one as rather not useful ($M = 3.5$), while two items on the trustworthiness of the explanations reached means of $4.2$ and $3.8$ with no rating below the midpoint. The single low usefulness rating came from the most experienced respondent and did not concern explanation quality. That participant noted that the motivation to start reviewing a thousand findings is often missing, and still gave the highest trust rating of all respondents because one trusts the underlying expert system. This matched the dominant theme of the discussions, namely the difficulty of dealing with findings at scale. Participants described the number of findings as overwhelming without support for grouping or prioritization, reported difficulty in assessing relevance, risk, and impact, and lacked a clear entry point once a scan produces hundreds of results.

Participants valued being able to trace the path of unsafe data through the code, described the step by step data flow view as the best aspect of the plugin, found the explanations far more understandable than the SAST output, and considered the IDE integration valuable because it places support in their working context. At the same time they reported that explanations still contain too much text and do not make the relevant information immediately visible, asking for colored highlighting, markers that indicate which part of the text answers which question, and a very short summary. Generic examples were seen as insufficient, since mapping such an example onto the concrete finding is exactly the work one would have to do without a tool. Participants therefore asked for quick fixes, mitigation strategies, and example code derived from their own project, together with estimates of how critical a finding is and how much effort a fix takes. On usability they asked for affected files to be marked in the file tree, for short explanations at the affected code, for a clearer link between an explanation and the underlying data flow, and for explicit source attribution.

\subsection{Takeaways}
\label{sec:takeaways}

The preliminary study told us how a single explanation should read. The workshop told us how explanations must behave once a developer meets hundreds of them inside a real project. We consolidate both into five takeaways that drive the requirements in Section~\ref{sec:requirements}.

\begin{itemize}
    \item[\textbf{T1}] \textbf{Conciseness is a matter of structure, not only of length.} The preliminary study identified verbosity as the main weakness, and although the second version already shortened the explanations, workshop participants still reported too much text. They asked for a visible hierarchy, a short summary at the top, and markers that reveal which part of the text answers which question, so conciseness has to come from progressive disclosure rather than from compression alone.
    \item[\textbf{T2}] \textbf{Actionability has to reach the user's own code.} Both studies rated mitigation guidance as the weakest part of the explanation, and the workshop sharpened the earlier request for corrected code examples into a criterion, since a generic example shifts the translation work back to the developer. Actionable support means fixes and examples derived from the analyzed project, complemented by an indication of severity and remediation effort.
    \item[\textbf{T3}] \textbf{The unit of work is the finding set, not the single finding.} This takeaway has no counterpart in the preliminary study, whose participants assessed one isolated warning at a time, whereas in practice the volume of findings dominates. The most experienced respondent framed missing prioritization as the reason a vulnerability explainer may still go unused, so explaining findings well while leaving their ordering to the user solves only part of the problem.
    \item[\textbf{T4}] \textbf{Explanations must be visibly anchored in the analysis and in the code.} The preliminary study observed explanations that framed false positives as real vulnerabilities and occasionally contradicted the SAST tool, which we attributed to reasoning over an isolated snippet, and the workshop confirmed the same gap from the opposite direction, as participants valued the data flow view yet could not always see how the explanation related to it. Explanations must therefore be traceable to the reasoning and evidence of the underlying analysis, which means to the rule whose conditions the analyser checked as well as to the data flow it reported, with explicit sources and affected locations marked in the editor and the file tree.
    \item[\textbf{T5}] \textbf{Trust and adoption depend on the surrounding workflow.} Trustworthiness was rated favorably, but the free text responses show that this trust was largely inherited from the established tools behind the plugin rather than earned by the explanation itself, and participants named fast results, low resource consumption, and integration into the environments and pipelines they already use as conditions for adoption. The preliminary study reached a related conclusion methodologically, since its screenshot-based setting could not capture live IDE interaction, so the explanation has to live where the work happens and be evaluated there as well.
\end{itemize}

Takeaways T1, T2, and T4 concern the explanation itself, while T3 and T5 concern the workflow into which it is embedded. The plugin should therefore not be understood as a text generation layer on top of static analysis results, since its usefulness depends on the broader socio-technical context in which developers, security experts, and product teams work with security findings.

\subsection{Requirements}
\label{sec:requirements}

Based on the takeaways T1 to T5 of Section~\ref{sec:takeaways}, we derive five key requirements for the second version of the SAFE plugin:

\begin{enumerate}
    \item[\textbf{R1}] \textbf{Conciseness and comprehensibility (T1).} The plugin should provide concise, comprehensible explanations accessible to users with different security expertise levels, communicating essential information without overwhelming detail.
    \item[\textbf{R2}] \textbf{Code and data-flow linkage (T4).} Explanations should be clearly linked to the underlying warning, the affected code location, and the relevant data flow.
    \item[\textbf{R3}] \textbf{Prioritization and actionability (T2, T3).} The plugin should support prioritization and actionability by helping users assess relevance, identify remediation starting points, and make informed decisions.
    \item[\textbf{R4}] \textbf{IDE-workflow integration (T5).} The plugin should integrate smoothly into existing workflows through contextual IDE visibility, code annotations, and file-tree markers.
    \item[\textbf{R5}] \textbf{Transparency and trust (T4, T5).} Explanations should be transparent and trustworthy through traceability, links to evidence, and appropriate communication of uncertainty.
\end{enumerate}

These requirements complement the general usability criteria for static analysis tools identified in prior work~\cite{10.1145/3533767.3534374, nachtigall2019explaining}, such as internal reasoning, clickable traces, explaining consequences, explanation by example, and connecting warnings to code. While those criteria address the warning message itself, R1--R5 additionally capture the workflow-level concerns raised by the workshop participants: managing and prioritizing large numbers of findings, acting on them within the IDE, and trusting the generated explanations. Section~\ref{sec:safe-agent} describes how the redesigned plugin addresses these requirements.

\section{Approach}\label{sec:safe-agent}

\begin{figure}[t]
\centering
\includegraphics[width=\linewidth]{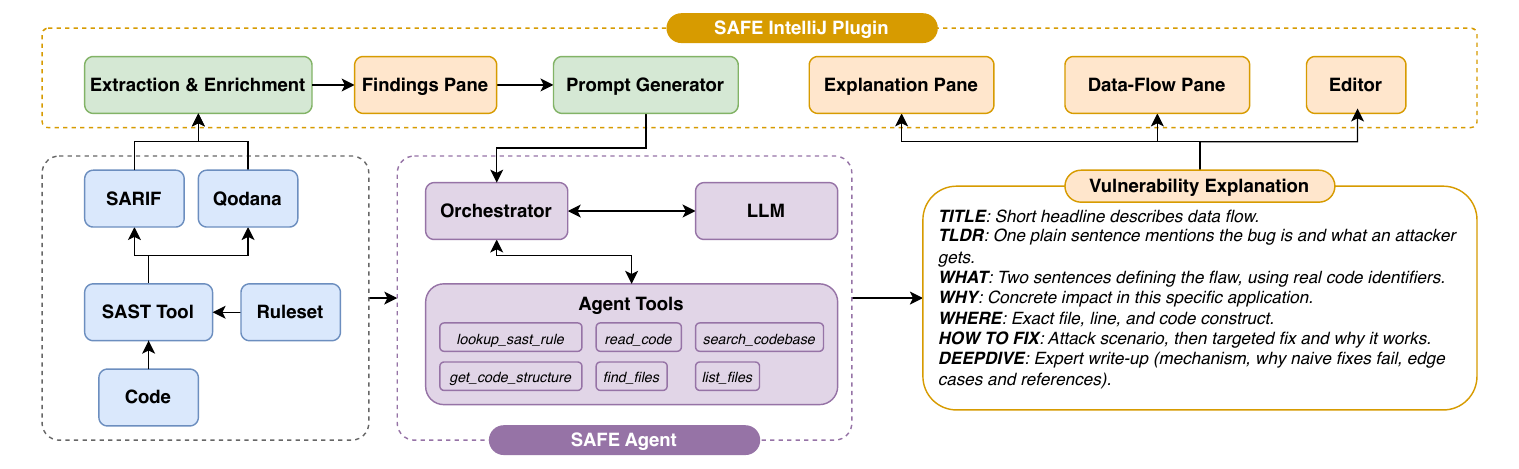}
\caption{Architecture of the SAFE approach. A SAST tool applies its ruleset to
the code and reports findings as SARIF or through Qodana. The SAFE plugin
extracts and enriches them, lists them in its findings pane, and builds a prompt
for the selected finding. In the SAFE Agent, an orchestrator relays that prompt
to the language model and executes the tool calls the model issues against the
project. The resulting structured explanation is rendered in the explanation
pane, the data-flow pane, and the editor.}
\label{fig:safe-approach}
\end{figure}

Guided by the requirements derived in Section~\ref{sec:requirements}, we redesigned the SAFE plugin and its explanation generation. SAFE deliberately does not detect vulnerabilities itself but instead makes the analyser's verdict intelligible to the developer who has to act on it. Detection remains with an established SAST tool that applies its own ruleset to the code, and whose results are auditable and reproducible. Figure~\ref{fig:safe-approach} shows the architecture, consisting of the \emph{SAFE plugin} and The \emph{SAFE Agent}. The \emph{SAFE plugin} extracts and enriches the findings inside IntelliJ IDEA, turns a selected finding into a prompt, and renders the returned explanation across the explanation and data-flow panes and the editor. The \emph{SAFE Agent} is a separate service in which an orchestrator drives a language model that must consult the user's codebase through a set of tools before it answers, so that the explanation is grounded in the project's source code rather than in generic textbook knowledge of the underlying weakness class. In the next sections, we describe the SAFE's architecture, the plugin's user interface, agentic grounding, and the structured explanations.

\subsection{Architecture and Design Rationale}
\label{sec:safe-architecture}

Findings enter the plugin through two routes. First, SAFE integrates with Qodana~\cite{Qodana}, the JetBrains SAST plugin for IntelliJ IDEs, and extracts the findings displayed in its tool window. Second, SAFE parses files in the Static Analysis Results Interchange Format (SARIF) 2.1.0~\cite{sarif}, so that the findings of any SAST tool that exports SARIF can be explained. Each finding is normalized to the rule identifier, severity, warning message, source location, the affected code snippet, and, when available, the source-to-sink data-flow trace.

That trace and the rule behind it are the analyser's own reasoning, and they are the starting point rather than something the model reconstructs. The agent is required to look up the rule that fired and to read the endpoints of that trace before it writes anything. An explanation is therefore an account of why this rule objected to this code path, anchored to the evidence the analyser produced, rather than an independent opinion about whether a vulnerability exists. This is what separates the approach from LLM-only detection, and it is why the plugin keeps the analyser's verbatim message beside the generated text instead of replacing it.

Each finding is enriched with deterministic metadata. SAFE resolves the inspection identifier to a Common Weakness Enumeration (CWE) entry~\cite{cwe} using a curated mapping together with a bundled copy of the MITRE CWE catalogue. This yields a ground-truth weakness name and description that both the prompt and the user interface reference. Resolving the CWE outside the LLM avoids misattributed weaknesses that would undermine the faithfulness of the explanation and the trust of its users (R5). In addition, the steps of the data-flow trace are classified into the roles \textsc{source}, \textsc{propagator}, \textsc{call}, \textsc{sanitizer}, and \textsc{sink}, which later makes individual steps interpretable in isolation.

Enriched findings are listed in the findings pane, and the finding the developer selects is handed to a prompt generator that assembles the request from the normalized fields, the resolved CWE, and the classified data-flow steps. The request is then passed to the active explanation provider. Alongside the SAFE Agent of Section~\ref{sec:safe-agent-backend}, which Figure~\ref{fig:safe-approach} depicts, the plugin also supports direct single-request providers, so that a deployment without the agent service still produces an explanation. The agent is the design we propose, and it is the backend used in the redesigned build we evaluate.

\subsection{Implementation}

The agent runs as a separate service in Python and FastAPI, built on LangChain and LangGraph~\cite{langchain, langgraph}, and streams partial results to the plugin over server-sent events so that a developer sees the explanation form rather than waiting on a complete response. Whether the SAFE agent or a single-request LLM provider is used is configured in
the plugin settings and the credentials are stored in the IDE's password safe.
Every backend returns an explanation conforming to the schema of Section~\ref{sec:safe-schema}, and generated explanations are cached persistently. Reopening a project therefore issues no new model calls, and the same finding can carry parallel explanations from different backends.

\subsection{SAFE Plugin User Interface}
\label{sec:safe-ui}

\begin{figure}[t]
\centering
\definecolor{annoFind}{HTML}{E87BA4}
\definecolor{annoExpl}{HTML}{008300}
\definecolor{annoFlow}{HTML}{4A3AA7}
\begin{tikzpicture}
\node[anchor=south west, inner sep=0] (img) at (0,0)
  {\includegraphics[width=\linewidth]{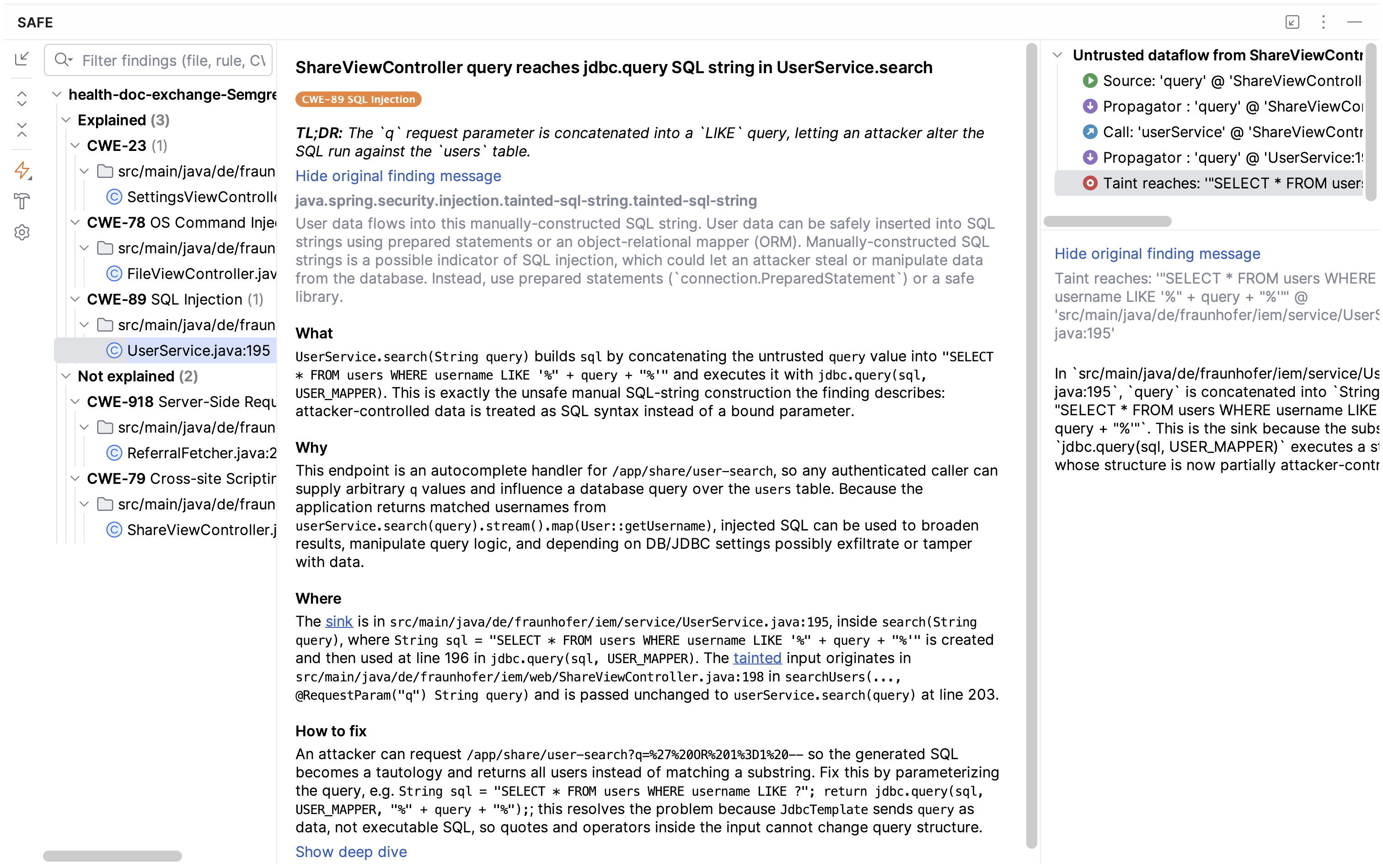}};
\begin{scope}[x={(img.south east)}, y={(img.north west)}]
  \tikzset{
    pane/.style={line width=0.9pt, rounded corners=1pt},
    plabel/.style={font=\bfseries\scriptsize, inner xsep=3pt, inner ysep=1pt,
                   fill=white, text opacity=1},
    badge/.style={circle, draw=annoExpl, fill=white, line width=0.5pt,
                  inner sep=0pt, minimum size=1.8mm,
                  font=\bfseries\tiny, text=annoExpl},
  }
  \draw[pane, draw=annoFind] (0.026,0.368) rectangle (0.196,0.951);
  \draw[pane, draw=annoExpl] (0.202,0.006) rectangle (0.7475,0.951);
  \draw[pane, draw=annoFlow] (0.7525,0.755) rectangle (0.9985,0.951);
  \draw[pane, draw=annoFlow] (0.7525,0.006) rectangle (0.9985,0.722);
  \node[plabel, anchor=south, text=annoFind] at (0.111,0.952) {Findings};
  \node[plabel, anchor=south, text=annoExpl] at (0.475,0.952) {Explanation};
  \node[plabel, anchor=south, text=annoFlow] at (0.855,0.952) {Data flow};
  \node[plabel, anchor=south, text=annoFlow] at (0.855,0.724) {Step explanation};
  \node[badge] at (0.2070,0.923) {1};
  \node[badge] at (0.2070,0.846) {2};
  \node[badge] at (0.2070,0.798) {3};
  \node[badge] at (0.2070,0.617) {4};
  \node[badge] at (0.2070,0.484) {5};
  \node[badge] at (0.2070,0.311) {6};
  \node[badge] at (0.2070,0.158) {7};
  \node[badge] at (0.2070,0.018) {8};
\end{scope}
\end{tikzpicture}
\caption{The redesigned SAFE tool window on a SQL-injection finding. \emph{Findings}
groups an analyser run by explanation status, CWE, directory and file, with a filter over
file, rule and CWE. \emph{Explanation} is numbered by region: \textbf{1}~title and CWE
badge, \textbf{2}~one-sentence TL;DR, \textbf{3}~a disclosure holding the analyser's own rule identifier and
message verbatim, shown expanded here, \textbf{4--7}~the \textsc{what}, \textsc{why}, \textsc{where} and \textsc{how to fix} sections, \textbf{8}~the opt-in deep dive.
\emph{Data flow} renders the trace as role-classified steps that each link to their file
and line. Selecting a step opens its own explanation below the list, above it the
verbatim step message from the analyser, here for the \textsc{sink} at the end of the
trace.}
\label{fig:plugin-ui}
\end{figure}

The workshop feedback (Section~\ref{sec:industry-workshop}) and the preliminary study (Section~\ref{sec:pre-study}) directly shaped a redesign of the plugin's user interface. Figure~\ref{fig:plugin-ui} shows the result on a SQL-injection finding reported by a SAST tool in a Java web application. The tool window is organised into three panes: findings, explanation and data-flow. \emph{Findings} shows which static analysis findings exist and which of them have already been explained. \emph{Explanation} gives information about the meaning of a selected finding. \emph{Data-flow} indicates how the untrusted data reaches the vulnerable operation. The numbered regions of the explanation pane are referenced individually below.

\subsubsection{Importing Findings}
A single action populates the tool
window either from the results the Qodana plugin is currently displaying or from a SARIF
file chosen through a file picker. The \emph{Not explained} branch
in Figure~\ref{fig:plugin-ui} holds findings that are already present, navigable, and
marked in the editor.
A dropdown in the action bar switches the active provider, marks the current one, and
refreshes the tree so that it shows which findings already carry an explanation from that
provider. The same control is what let a single finding hold parallel explanations from
different backends during the evaluation in Section~\ref{sec:evalaution}.

\subsubsection{Finding Tree}
The findings pane groups the findings of an analyser run by explanation status, CWE, directory, and file, with severity icons and a CWE badge whose tooltip shows the catalogue description, and a filter narrows the tree by file, rule, or CWE. The figure shows three findings already explained and two not yet examined. This gives an overview of \emph{what kinds} of weaknesses the analyser reported and which of them have already been examined. It supports prioritization across large numbers of findings, which was one of the dominant pain points named in the workshop (R3).

\subsubsection{Tiered Explanations}
The first version tiered explanations by \emph{audience}, using a selector that regenerated the entire explanation at a user-chosen depth of beginner, intermediate, or advanced. Participants nevertheless found the explanations too long and asked for key information to be visible at a glance. The redesigned interface therefore tiers by demand. Every finding opens with a title and its CWE badge (1) and a one-sentence TL;DR (2), followed by the what (4), why (5), where (6) and fix (7) core sections, while the in-depth material sits behind the opt-in deep-dive disclosure (8). The analyser's own rule identifier and message are preserved verbatim behind a further disclosure (3), shown expanded in the figure, so the rewriting by the LLM never hides the verdict it is rewriting (R1, R5). An inline glossary annotates security terms such as \emph{taint} and \emph{deserialization} with hover definitions, which keeps the default text concise without losing precision (R1).

\subsubsection{Data-flow Visualization}
When the finding carries a taint trace, the data-flow pane renders it as a stepped list in which each hop is labelled with its role and links to its file and line. Every step also carries an explanation of its own. Selecting a step opens it below the list, as Figure~\ref{fig:plugin-ui} shows for the \textsc{sink} at the end of the trace. These per-step explanations are requested as part of the same structured response as the rest of the explanation (Section~\ref{sec:safe-schema}), so opening one costs no further LLM call and the trace stays consistent with the prose that describes it. Each one says what that particular hop does to the untrusted value at that file and line, which is the question the raw trace leaves open. The analyser's own message for the step is kept above it verbatim behind the same kind of disclosure used for the finding message (3), so a developer can always compare the generated wording against what the tool reported (R5). Because a step is only interpretable once its role is known, this reuses the deterministic classification into \textsc{source}, \textsc{propagator}, \textsc{call}, \textsc{sanitizer}, and \textsc{sink} rather than asking the LLM to infer it. 

Opening a step explanation and jumping to the corresponding editor location remain separate actions, so the same progressive disclosure that governs the explanation text also governs the trace (R1). The \textsc{where} section (6) links back to the same source and sink locations, so the prose and the trace refer to each other instead of sitting side by side. Workshop participants valued the flow view most but could not always see how it related to the explanation text, and these links close that gap (R2).

\subsubsection{Workflow Integration}
The tool window is only one of the two interfaces the plugin uses. The other is the editor, which is where developers spend their time and where workshop participants wanted findings to arrive instead of in a separate dashboard (R4). Every imported finding receives a gutter icon and a highlight on the flagged line at import time, so an analyser run is visible in the code before any explanation has been requested. Hovering a gutter icon shows the explanation for that line in a tooltip, which makes a quick check possible without opening the tool window at all. When a finding carries a taint trace, the individual steps are highlighted as well, each in the colour of its role, so a developer can follow the untrusted value through the code in the same terms the data-flow pane uses. Navigation runs in both directions. Selecting a finding or one of its steps moves the caret to the corresponding location, and the gutter icon opens the matching explanation in the tool window. Because explanations are cached persistently, reopening a project restores both the tree and the highlights without issuing new LLM calls. An editor listener reapplies the highlights whenever a file with stored explanations is opened again. An explanation therefore stays attached to the code it describes instead of living in a report that the developer has to keep in a second window.

\subsection{Structured Explanations}
\label{sec:safe-schema}

Every explanation backend must return its answer in a fixed schema. Its sections decompose the explanation along the questions developers ask when diagnosing potential vulnerabilities with static analysis tools~\cite{question_dev_ask}. We adopt in particular the \emph{what}, \emph{where}, \emph{why}, and \emph{how} decomposition of SECGen~\cite{secgen}, which restructures raw SAST output into a developer-friendly report and thereby also admits automated validation of the generated text. Answering these four questions for every finding is what makes the explanation comprehensible to users with different levels of security expertise (R1). Table~\ref{tab:schema} lists the resulting sections and what each of them must contain.

\begin{table}[t]
\caption{The explanation schema that every explanation backend must return, which the plugin parses and renders independently.}
\label{tab:schema}
\begin{tabular}{@{}p{0.14\linewidth}p{0.80\linewidth}@{}}
\toprule
Section & Required content \\
\midrule
\textsc{title}    & A short headline naming the data flow behind the finding. \\
\textsc{tldr}     & One plain sentence stating what the reported vulnerability is and what an attacker gains from it. \\
\textsc{what}     & Two sentences defining the vulnerability in terms of the real code identifiers involved rather than of the rule identifier that fired. \\
\textsc{why}      & The concrete impact in this specific application rather than in the weakness class in general. \\
\textsc{where}    & The exact file, line, and code construct at which the problem manifests. \\
\textsc{how to fix}      & An attack scenario, followed by a targeted fix and an account of why that fix works. \\
\textsc{deepdive} & Optional expert write-up of the mechanism, of why naive fixes fail, and of edge cases and references. Rendered behind the opt-in disclosure described above. \\
\textsc{step}     & One short annotation per hop of the taint trace, returned only when the finding carries one and rendered in the data-flow pane. \\
\bottomrule
\end{tabular}
\end{table}

Fixing the structure serves three purposes. The plugin can parse and render each section independently in its tiered layout, and it can check automatically that a backend returned every required section before the explanation reaches the user. The developer encounters every explanation under the same headings, which builds a stable mental template and reduces orientation effort over time (R1).  Finally, explanations produced by different backends remain structurally comparable.

\subsection{Code-Grounded Explanations}
\label{sec:safe-agent-backend}

A direct LLM call is fast, but it can only describe the weakness class in general terms, because the model sees the warning message and a code snippet rather than the project the finding lives in. The industry workshop surfaced the consequences. Participants found the resulting explanations verbose, hard to connect to their own code, and prone to arguing that a false positive is a real vulnerability. The SAFE Agent addresses this by \emph{grounding} the explanation in the user's actual codebase. The same finding, explained by the agent, references the developer's real variables, methods, and call sites, and reflects what the triggered rule actually checks.

The plugin treats the agent as just another explanation provider, sending the same prompt and receiving the same explanation schema, so users switch between agentic and direct explanations from the same control. Inside the agent, an orchestrator relays the prompt to the language model, executes the tool calls the model issues against the project, and feeds their results back, repeating this until the model returns the final explanation. The model must therefore consult the repository before answering. Table~\ref{tab:tools} lists the six tools available to it. Two of them establish what the analyser checked and where, two retrieve code, and two navigate the project tree.

\begin{table}[t]
\caption{The tools available to the SAFE Agent for evaluating the SAST rules, source code and project tree.}
\label{tab:tools}
\begin{tabular}{@{}p{0.27\linewidth}p{0.67\linewidth}@{}}
\toprule
Tool & Purpose \\
\midrule
\texttt{lookup\_sast\_rule}   & Looks up the definition of the rule that fired and returns its identifier, message, patterns, severity, and CWE and OWASP metadata, so that the explanation reflects what the rule actually checks. \\
\texttt{get\_code\_structure} & Returns the functions, classes, and methods of a file with their exact line ranges, taken from a precomputed index. \\
\texttt{read\_code}           & Reads source lines from a file, optionally restricted to a line range. \\
\texttt{search\_codebase}     & Searches the codebase for a pattern, optionally restricted to a list of files, and returns the matching paths. \\
\texttt{list\_files}          & Lists the entries of a directory, with the output truncated to a bounded number of entries. \\
\texttt{find\_files}          & Finds files whose name or path contains a given substring. \\
\bottomrule
\end{tabular}
\end{table}

The system prompt mandates that the agent first looks up the rule that fired and reads the flagged code span together with the endpoints of the data-flow trace. It explicitly forbids generic textbook responses that do not reference the inspected code. The rule lookup matters because SAST rules are only noisy proxies for the CWE they are tagged with. The same CWE can be reported by structurally different rules, and a rule frequently encodes assumptions that a CWE-level description would miss. Grounding the explanation in what the rule actually checks aligns it with the analyser's verdict rather than with a generic description of the weakness class (R2).  

This is a narrower remit than a general-purpose coding agent carries~\cite{3737916.3739517, 3692070.3694124, aider}. The agent produces an account a developer reads rather than a patch, stays bound to a rule's existing verdict rather than forming one, and navigates only the artefacts that substantiate the reported trace. It can neither modify code nor issue a verdict, which is why the tool set in Table~\ref{tab:tools} is deliberately small.

Tool use also changes what the developer can observe. The stream carries an event for every tool invocation, so the sequence of consulted files and rules constitutes an evidence trail, so each explanation is traceable to the concrete artifacts the agent read (R5). Finally, on-demand file access scales to repositories of arbitrary size, since the agent retrieves only the artifacts it decides to consult instead of filling the context window of the model with the codebase.

\section{Evaluation}\label{sec:evalaution}

To assess whether the redesign of the SAFE plugin improves triage support in practice, we conducted a controlled within-subjects user study that follows the ACM SIGSOFT Empirical Standard for experiments with human participants \cite{10.1145/3437479.3437483}. 
The study evaluates the first version of the SAFE plugin against the redesigned plugin. Throughout the study, the plugins are identified using deliberately neutral names, and we retain them here. \emph{SAFE-Blue} is the first version of the plugin which explains a vulnerability with a single direct LLM request and tiers the result by audience through a beginner, intermediate, and advanced depth selector.
\emph{SAFE-Red} is the second version plugin, which replaces that selector with the fixed tiered presentation of Section~\ref{sec:safe-ui} and generates its explanations with the agent of Section~\ref{sec:safe-agent-backend}. The colours carry no meaning beyond identifying the two builds to participants.

\subsection{Research Questions and Hypotheses}
\label{sec:eval-rqs}

We ask three research questions about SAFE's redesign as a whole, because that is what participants used and what the study can speak to. SAFE-Red bundles the tiered presentation of Section~\ref{sec:safe-ui} (R1), the code-grounded agentic explanations of Section~\ref{sec:safe-agent-backend} (R2, R3), and the workflow integration of Section~\ref{sec:safe-ui} (R4, R5). These are not separable in the artefact and are not separated by the design, so the questions name the system. The study establishes whether a redesign built to satisfy R1 to R5 performs differently from the design it replaces.

\begin{enumerate}
    \item[\textbf{RQ1}] Does the requirements-driven redesign of SAFE, comprising structured tiered presentation, code-grounded agentic explanations, and IDE-workflow integration, change participants' decision quality relative to the prior single-prompt design, both for detection and for remediation?
    \item[\textbf{RQ2}] Does the redesigned plugin change time-to-decision relative to the prior design?
    \item[\textbf{RQ3}] Do participants rate the redesigned plugin differently from the prior design on explanation quality, IDE-workflow integration, usability, and trust?
\end{enumerate}

RQ1 is split into H1a on triage correctness and H1b on fix-family correctness, because R2 and R3 attach to remediation independently of detection. RQ2 covers H2 on time per finding. RQ3 registers one hypothesis per perception construct named by a requirement rather than a single aggregate, namely H3a on perceived explanation quality with its clarity, actionability, and confidence-support sub-constructs (R1, R3, R5), H3b on usability, H3c on IDE-workflow integration (R4), and H3d on trust (R5).

Every hypothesis is stated in null and alternative form and every alternative is two-sided. We decline a directional commitment because no member of the family has unambiguous prior evidence. The default SAFE-Red explanation is shorter and more structured than the SAFE-Blue one, which could equally help by removing noise or hurt by placing essential detail behind an opt-in tier. Its optional deep-dive layer resembles the SAFE-Blue explanation and could equally help through deeper grounding or hurt by inducing over-reliance among the participants who open it. The same reasoning applies to time, since a shorter default may speed decisions while an unfamiliar interface and an opt-in layer may slow them. The family-wise significance level is $\alpha = .05$.



\subsection{Design and Procedure}
\label{sec:eval-design}

The experiment is a two-condition AB/BA crossover in which both order and snippet set are crossed with tool, giving the four sequences in Table~\ref{tab:sequences}. Prior to the sessions, assignment of participants to the four groups was randomised and stratified.
Across the four groups every combination of tool and set is observed equally often and every set appears equally often in each period. This dissolves the object-treatment confound that would otherwise arise from pairing one tool with one set. A within-subjects design was chosen because it lets each participant serve as their own control on personal-skill variance. A between-subjects design, by contrast, would have required roughly twice as many
participants to reach comparable statistical power, which exceeded the pool available to us. The cost is potential carry-over, which the counterbalancing above is intended to absorb.

\begin{table}[t]
\caption{The four counterbalanced sequences. Each group crosses tool order with snippet-set assignment.}
\label{tab:sequences}
\begin{tabular}{@{}lll@{}}
\toprule
Group & Task Block 1 & Task Block 2 \\
\midrule
G1 & SAFE-Blue on Set 1 & SAFE-Red on Set 2 \\
G2 & SAFE-Red on Set 2  & SAFE-Blue on Set 1 \\
G3 & SAFE-Blue on Set 2 & SAFE-Red on Set 1 \\
G4 & SAFE-Red on Set 1  & SAFE-Blue on Set 2 \\
\bottomrule
\end{tabular}
\end{table}

The session were estimated to last for 60 minutes and followed the outline below: 

\begin{enumerate}
  \item \textbf{Consent:} Participants consented in advance and again verbally at the start, both to the session recording, which was used to validate the transcript and then deleted, and to the logging of their interactions with the plugin.
  \item \textbf{Pre-study questionnaire:} Recording the participant's primary role, years of professional development experience, years of Java experience, frequency of JetBrains IDE use, self-rated familiarity with secure coding practices, and prior exposure to SAST tools.
  \item \textbf{Warm-up:} A video on a single fixed deserialization finding (CWE-502) with no plugin present. It familiarised participants with the IntelliJ IDEA interface and software project before either plugin was introduced, so that first-contact effects did not land on whichever tool came first.
  \item \textbf{Two task blocks:} In the order given by the participant's group in Table \ref{tab:sequences}, each block ran through the same four steps.
    \begin{enumerate}
      \item \textbf{Tool demo:} A video for the tool of that period. Both demos follow the same structural template, so neither condition received the richer introduction.
      \item \textbf{Setup:} The experimenter enabled the interaction telemetry and entered the participant identifier. Explanations for the assigned set had been generated beforehand and written to the plugin's cache, using a study-only action that walks every imported finding and requests an explanation from the active provider. Prewarming the cache means that navigating the tree never waited on the model, so the measured time reflects reading and deciding rather than generation latency, at the cost of removing generation latency from the comparison altogether.
      \item \textbf{Task:} Fifteen minutes to work through the five findings of the assigned set with the assigned tool. For each finding, participants decided whether it was a true positive, a false positive, or that they were unsure, described or sketched a fix, and recorded their confidence. They were told explicitly that finishing all five was not expected, which keeps the time limit from turning the task into a completion race. Think-aloud was encouraged, and the experimenter intervened only with neutral cues such as asking what the participant was looking at.
      \item \textbf{Block questionnaire:} Administered immediately, so that perceptions were recorded while the condition was still fresh rather than after both had been seen.
    \end{enumerate}
  \item \textbf{Interview and debrief:} A semi-structured interview covering what participants liked and disliked about the tool they had just used, what they would have done differently with the other one, what their trust in the assessment rested on, and how they would explain a finding to a colleague, followed by a debrief.
\end{enumerate}

\subsection{Subject Application and Findings}
\label{sec:eval-materials}

The findings come from the Health-Doc-Exchange project, a Java web application whose size and structure resemble a real project and for which a curated ground truth is available. The application is part of the material used to train security champions in the software security trainings offered by Fraunhofer IEM~\cite{iem-academy}. Those trainings were themselves developed in response to a study that found developers, product owners, and managers to have too little competence in secure software development \cite{dziwok2021appsecure}. The application therefore predates this experiment, and its weaknesses were curated for teaching rather than chosen to suit either condition. 

Choosing an internal application over a public benchmark was itself deliberate. Health-Doc-Exchange is not published, so its source is very unlikely to appear in the pretraining data of either model, and an explanation of one of its findings has to be derived from the code supplied at inference time rather than recalled from the code the model was trained on. A public corpus offers no such assurance, including the OWASP Benchmark \cite{owasp_benchmark_git} used in the preliminary study of Section~\ref{sec:pre-study} and the first version of the SAFE plugin.
Using an existing application rather than synthetic snippets matters for the agentic condition in particular, since an agent that searches the repository can only be exercised by code with genuine structure to search.

We scanned the application with Semgrep v1.160.0 and retained findings that carry at least one valid CWE mapping, spread across five weakness families relevant to back-end Java, namely path traversal (CWE-23), OS command injection (CWE-78), cross-site scripting (CWE-79), SQL injection (CWE-89), and server-side request forgery (CWE-918).

The retained findings are split into the two sets of Table~\ref{tab:sequences}, each holding one finding per weakness family, with a mix of true and false positives so that dismissing every warning is not a winning strategy. The sets were balanced ahead of time on CWE distribution, on the intended ratio of true to false positives, and on snippet length and cyclomatic complexity. The realised composition is two true positives and three false positives in one set against three and two in the other.

\subsection{Results}
\label{sec:eval-results}

Twenty-four participants participated in the study, distributed equally over the four sequences. Among the participants were software developers (10), computer science students (4), researchers (2), and a security engineer (1). Seven participants indicated another role which included two adjacent technical roles (software architect and release engineer), management roles outside of software development (2), two unrecorded roles, and a product team member. Fifteen had three or more years of professional development experience, and Java experience skewed lower, with twelve reporting less than a year. Fourteen had never used a SAST tool, and sixteen had remediated a vulnerability at least occasionally. Self-rated familiarity with secure coding clustered at the lower middle of the five-point scale.

One correction was applied after data collection. The recording for one participant shows SAFE-Red used first, while the response form recorded the assigned order of SAFE-Blue first, so we relabelled that participant's two blocks. All 24 sessions were checked by the same criterion, the order in which each tool was introduced, and this was the only disagreement. The relabelling leaves untouched which findings the participant saw, so their verdicts stand as graded, but it moves them to another sequence. The four groups therefore hold six, seven, five and six participants rather than six each, and each tool is paired with one finding set thirteen times and the other eleven. Each set still appears twelve times per period, so order and carry-over protection is unaffected.

\subsubsection{Task Performance: Verdicts and Remediations (RQ1)}

RQ1 asks whether the redesign changes the quality of the decisions participants reach, and it carries two hypotheses. H1a concerns triage correctness, the verdict of true or false positive measured against the ground truth. H1b concerns the correctness of the remediation participants proposed, which R2 and R3 attach to independently of detection. 

\paragraph{Triage Correctness (H1a)}
Participants recorded 124 of the 240 possible decisions. Counting an unsure verdict as an incorrect decision, per-participant accuracy averaged $0.46$ with SAFE-Blue and $0.53$ with SAFE-Red. The difference is not significant (Wilcoxon signed-rank, $n = 20$ non-tied pairs, $z = -0.54$, $p = .59$, rank-biserial $r = 0.14$). Eleven participants did better with SAFE-Red and nine with SAFE-Blue.

\begin{figure}[t]
\centering
\includegraphics[width=0.62\linewidth]{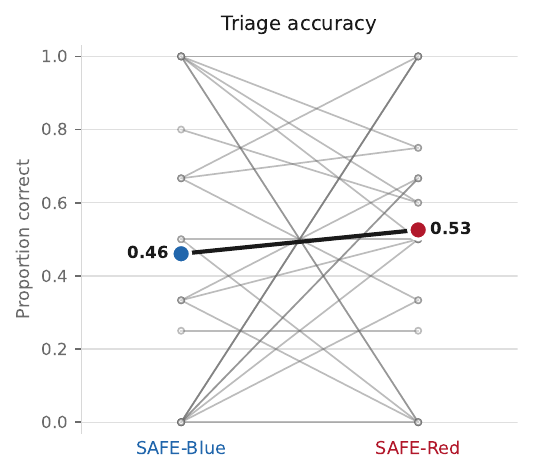}
\caption{Per-participant triage accuracy for the two conditions ($N=24$). Each thin line
joins one participant's two blocks, and the heavy line joins the condition means. An unsure
verdict counts as incorrect.}
\label{fig:results-accuracy}
\end{figure}

Excluding unsure verdicts instead of counting them as errors moves both conditions upward without separating them, to $0.56$ and $0.63$ over all decisions recorded in a condition. Figure~\ref{fig:results-accuracy} shows the per-participant pairs, and the crossing lines are the result. The two tools are indistinguishable on decision quality, and individual differences dominate the tool effect. Accuracy varied far more by the material than by the tool. The two finding sets differ by more than the tools do, at $0.41$ against $0.64$ across both conditions over 61 and 63 decisions, and the per-weakness cells vary more sharply still. We do not test either, since the per-weakness cells hold between four and twenty-one decisions, but the size of the gap is the clearest single reason to read the tool comparison on accuracy as uninformative.

\paragraph{Fix-family Correctness (H1b)}

The recording form captured each proposed fix as free text rather than as a choice from the controlled vocabulary, so the proposals were coded after the fact. Half of them are missing from the form, so we take the written response where a participant gave one and the proposal they made aloud, recovered from the session transcript, where they did not. The two sources are used in 33 and 32 cases and are close to evenly spread across the conditions, at 54 and 49 per cent written for SAFE-Blue and SAFE-Red. Coding covers the 64 findings that the ground truth marks as true positives, since a fix family is only defined where there is something to fix.

Two coders worked from a shared codebook, blind to participant and condition. They calibrated on 15 items, revised the codebook, and then coded the remainder with 20 further items in common. Agreement on the 20 post-revision items was 80 per cent exact, with a quadratic-weighted Cohen's kappa of $0.79$. The eight disagreements were settled by discussion, two of them by a rule the codebook already contained.

Proposals were rated correct for 19 per cent of the SAFE-Blue findings and 41 per cent of the SAFE-Red ones. We fit a cumulative-link model to the four-level rating, with standard errors clustered by participant. It puts the odds of a better-quality proposal with SAFE-Red at 2.6 times those with SAFE-Blue (95 percent CI 0.93 to 7.09, p = .069). Adding period and finding set as covariates raises the estimate to 2.8 (p = .064). The registered fallback, a Wilcoxon signed-rank test on per-participant means over the 15 participants who contributed findings in both conditions, agrees at $p = .068$. This is the largest separation between the two tools that the study produces on a performance measure, and it is the one that comes closest to significance. It also points the same way as the perception results, where actionability of the next steps was the widest gap. 

\finding{\researchq{1} Summary:}{Triage accuracy is indistinguishable between the two builds, at $0.46$ against $0.53$ ($p = .59$), and differences between participants outweigh the difference between tools. Remediation quality separates them. Nineteen per cent of proposals were rated correct with SAFE-Blue against 41 per cent with SAFE-Red, an odds ratio of $2.6$ ($p = .069$).}

\subsubsection{Task Completion Time (RQ2)}
RQ2 asks whether the redesign changes how long a finding takes. It carries H2 on time per finding.

Time per finding was not captured in the recording form and is not recoverable from the interaction telemetry, which logs how long a finding stayed selected rather than how long it took to decide. We therefore recovered it from the session transcripts, marking for each finding the moment the participant turned to it and the moment they first stated a verdict. This yields an open-to-decision interval that excludes the clerical time spent filling in the response form. It is available for 122 of the 124 recorded decisions, covering all 24 participants.

\begin{figure}[t]
\centering
\begin{subfigure}[t]{0.48\linewidth}
  \centering
  \includegraphics[width=\linewidth]{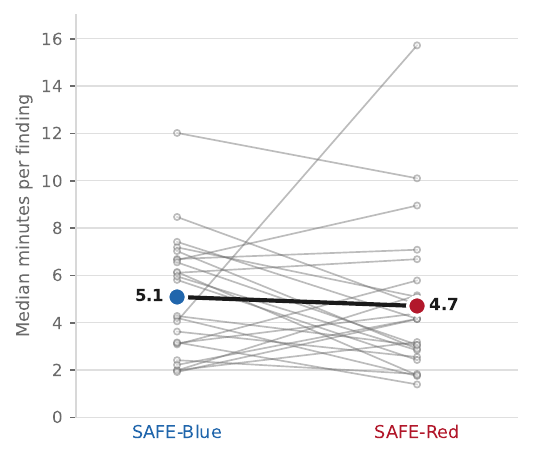}
  \caption{Per-participant medians for the two conditions.}
  \label{fig:time-paired}
\end{subfigure}\hfill
\begin{subfigure}[t]{0.48\linewidth}
  \centering
  \includegraphics[width=\linewidth]{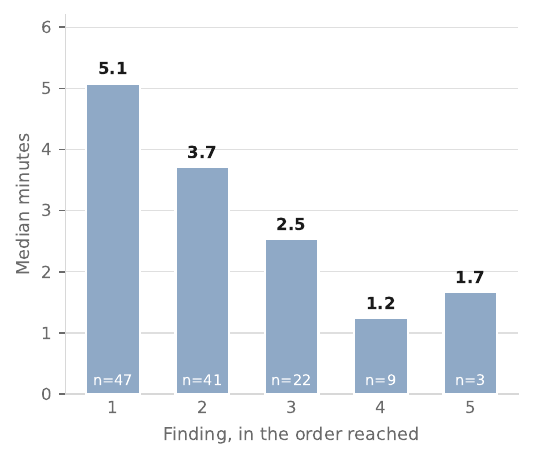}
  \caption{Median time by the position of a finding within its block, with the number of observations in each bar.}
  \label{fig:time-position}
\end{subfigure}
\caption{Time per finding, recovered from the session transcripts as the interval from the
participant turning to a finding to their first stated verdict.}
\label{fig:results-time}
\end{figure}

Participants took a median of $3.5$ minutes per finding. Per-participant medians average $5.1$ minutes with SAFE-Blue against $4.7$ minutes with SAFE-Red, a ratio of geometric means of $0.87$ in favour of SAFE-Red, and the difference is not significant (Wilcoxon signed-rank, $n = 24$, $z = -1.24$, $p = .21$, rank-biserial $r = -0.29$). Fourteen of the 24 participants were faster with SAFE-Red. Figure~\ref{fig:time-paired} shows the pairs. Efficiency therefore behaves like accuracy, with a direction that favours the redesign and a difference the study cannot distinguish from zero.

Figure~\ref{fig:time-position} explains why so few findings were reached. Time falls steeply with position in the block, from a median of $5.1$ minutes on the first finding to $3.7$, $2.5$ and $1.2$ on those that follow, so the first finding of a block costs $1.8$ times what a later one does. At the observed overall median, five findings would need $17.7$ minutes of decision time alone inside a fifteen-minute block, before any form filling. The participants who reached only two or three findings were not unusually slow, they were working at the rate the task demanded. This also bounds what the null means, since with a median of two findings per block roughly half of the timed material is first-contact orientation rather than steady-state triage.

\finding{\researchq{2} Summary:}{Time per finding is indistinguishable between the two builds, at a per-participant median of $5.1$ minutes against $4.7$ ($p = .21$), with 14 of the 24 participants faster with SAFE-Red. At the observed median of $3.5$ minutes per finding, five findings need more decision time than a fifteen-minute block contains, which is why most participants reached two or three of them.}

\subsubsection{Perceived Usability, Explanation Quality and Trust (RQ3)}
\label{sec:rq3}

RQ3 asks whether participants rate the two builds differently, and it registers one hypothesis per construct named by a requirement rather than a single aggregate. H3a covers perceived explanation quality with its clarity, actionability, and confidence-support sub-constructs, H3b usability, H3c IDE-workflow integration, and H3d trust.

\begin{figure}[t]
\centering
\includegraphics[width=0.62\linewidth]{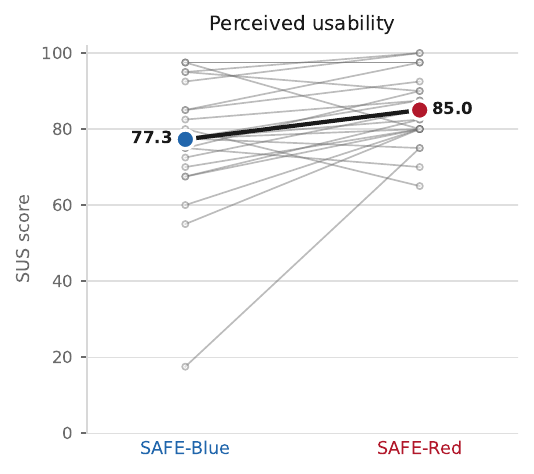}
\caption{Per-participant System Usability Scale scores for the two conditions ($N=24$).
Each thin line joins one participant's two blocks, and the heavy line joins the condition means.}
\label{fig:results-sus}
\end{figure}

\begin{figure}[t]
\centering
\includegraphics[width=\linewidth]{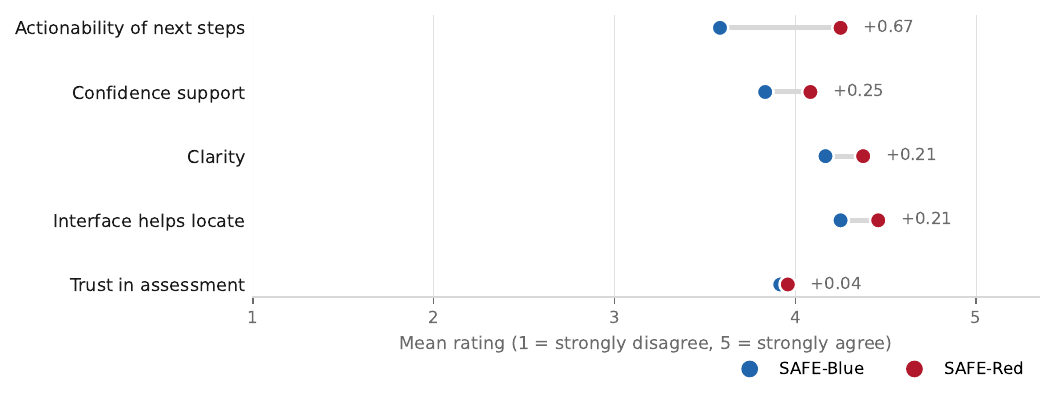}
\caption{Mean ratings on the five post-task perception items, ordered by the size
of the gap between conditions.}
\label{fig:results-perception}
\end{figure}

\paragraph{Usability, explanation quality and trust (H3a to H3d).}
SAFE-Red scored $85.0$ on the System Usability Scale \cite{sus} against $77.3$ for SAFE-Blue, with seventeen of the twenty-two non-tied participants rating it higher ($z = -2.55$, $p = .011$, $r = 0.62$), shown in Figure~\ref{fig:results-sus}. Figure~\ref{fig:results-perception} summarises the five perception items. SAFE-Red is rated at least as highly on every one, and the ordering of the gaps is itself informative. The largest is actionability of the next steps, at $+0.67$ on the five-point scale ($z = -2.53$, raw $p = .011$, $r = 0.73$), followed by confidence support at $+0.25$ and clarity and the interface helping to locate issues, both at $+0.21$. Trust in the assessment is effectively unchanged at $+0.04$. After Holm correction across the five items no individual comparison remains significant, actionability reaching $p = .057$. Willingness to recommend the plugin moved from $7.96$ to $8.67$ on the eleven-point scale and does not separate the conditions at the five per cent level ($p = .07$).

\paragraph{What participants actually did.}
The questionnaire records what participants said about each build. The interaction telemetry records what they did with it, and the protocol registers it for exactly this purpose, as a check on the self-reported ratings rather than as a hypothesis of its own. Telemetry maps to 36 sessions across 18 participants, 15 of whom have logs for both task blocks. The two plugins are not instrumented symmetrically, so we report usage within each condition rather than as a paired contrast. Two observations bear on the redesign. First, the opt-in deep dive was opened 12 times by 6 of 16 participants, so the concise default answered most findings on its own and the deeper layer was reached when it was wanted rather than routinely (R1). Second, the links that tie the explanation to the trace were exercised rather than ignored, with per-step explanations opened 134 times by 12 of 16 participants, jumps from a step to the editor taken 106 times by 11 of them, and glossary definitions shown 149 times by 12 (R2). The comparable behaviour in SAFE-Blue is clicking a data-flow step, which 10 of 17 participants did 152 times. These counts describe what participants did with each build and do not license a claim that one caused better outcomes.

\finding{\researchq{3} Summary:}{SAFE-Red is rated at least as highly on every item. Usability is the only comparison to reach significance, at $77.3$ against $85.0$ on the System Usability Scale ($p = .011$). Actionability shows the widest gap among the explanation items at $+0.67$, the construct both formative studies had identified as the weakest, although no individual item survives Holm correction.}

\subsubsection{Semi-Structured Interview Themes}\label{sec:eval-interviews}
The closing interviews were coded to explain the quantitative results rather than to test a hypothesis. Two coders developed a code system together on five transcripts, then applied it independently, with four interviews double-coded. Segment-level agreement across those four is $46.9$ per cent over 64 compared codings. Agreement divides sharply by code type. Codes with a concrete referent reach $66.7$ per cent, namely the automation boundary, the importance of explaining a vulnerability, data-flow navigation, code annotations and expertise level. Codes about presentation reach zero, namely explanation structure, clarity, text length and the interface. Presentation talk runs through every answer, so the two coders marked overlapping spans that the overlap criterion counts as disagreements. We therefore treat the themes below as an interpretive aid rather than as findings in their own right, and we report participant counts rather than testing anything. Coding covers 211 segments across all 24 interviews.

\paragraph{The automation boundary.} The most widespread theme is one that no requirement anticipated. Participants spoke to it in 22 of the 24 interviews, and 20 of them placed the final judgement with the developer. The tool should find and explain, and a developer should judge and apply. Four reasons recur. Six participants wanted to retain the knowledge that reviewing a finding gives them. Five wanted to see the extent of a change before it was applied, refusing an assistant that commits or merges on its own. Five cited the analyser's false positives, one warning that an automatically applied fix for a false positive introduces the defect it was meant to remove. Three treated the code as theirs to answer for, one saying the assistant will not be the one held responsible and another that the more a developer relies on the tool, the more the code becomes the tool's rather than their own. No participant argued for removing the developer, although three allowed that automation could go further given more confidence in the model or given time. This bears on the design commitment in Section~\ref{sec:safe-agent} rather than on any hypothesis, since the agent is constrained to explain a verdict it may not form, and the interviews indicate that developers want that boundary drawn where we drew it.

\paragraph{What trust rested on.} The interviews qualify the flat trust rating discussed in Section~\ref{sec:rq3}. Twelve participants discussed the data-flow navigation and twelve the code annotations. Six described these as raising confidence, saying that navigating the trace increased their confidence in the verdict. They mentioned that seeing the flagged line gave assurance the tool knew what it was talking about, and that an explanation anchored in a real source was worth more than one from a model treated as a black box. Two separated the two ideas explicitly, saying the annotations made the work easier without making the assessment more believable. A third made trust conditional on difficulty, trusting the output on a straightforward finding while expecting a harder one to be answered with something invented. Grounding therefore appears to support locating and understanding a finding rather than believing a verdict, which is consistent with actionability moving while trust did not.

\paragraph{Attributing the difference.} Explanation structure and content is the second most widespread theme, raised in 21 of the 24 interviews. Participants attributed the difference between the builds to how explanations were organised rather than to how they were produced. They highlighted that the fixed sections, the short summary, the opt-in depth and the reduced clicking were key factors. Several said that a build carrying the other's presentation would have changed which one they preferred. A further 17 discussed the expertise level and divided over whether such a control is wanted at all, which is one of the main design changes. 

\section{Discussion}\label{sec:discussion}



SAFE's redesign changed what participants thought of the tool and what they proposed as a fix, without changing the verdicts they reached. Every perception measure favours SAFE-Red, and usability significantly so. The widest gap among the explanation items is actionability. That is the construct prioritization and actionability (R3) targets, and the aspect both formative studies had identified as the weakest part of the earlier explanations (T2). Remediation quality moves with it, with proposals rated correct roughly twice as often. Actionability and fix quality are the same claim measured two ways, once by asking participants and once by grading what they wrote, and they agree. That convergence is the strongest evidence the study offers, because a rating alone would be open to the objection that a nicer-looking explanation simply feels more useful.

Triage accuracy does not follow, and neither does time. This is consistent with the ceiling the task imposed rather than with the explanations being unhelpful. At the observed median of $3.5$ minutes per finding, five findings need more decision time than a fifteen-minute block contains. Reaching two or three of them is therefore the expected outcome rather than a sign of slow work. No measure of decision quality separates two tools on two or three decisions per participant. We read the accuracy result as uninformative rather than as evidence of equivalence.

Read against the requirements, the pattern is uneven in an informative way. Conciseness (R1) holds, since the concise default answered most findings and the opt-in tier was opened rarely rather than routinely, which is what tiering by demand was meant to achieve. Code and data-flow linkage (R2) holds behaviourally, since the links from explanation to trace were exercised by most participants who had them. Actionability (R3) shows the largest movement of any construct. Transparency and trust (R5) does not move at all. It was the one perception item that stayed flat while everything around it rose. Grounding an explanation in the developer's own code evidently did not make participants more willing to believe the analyser's verdict. That is worth stating plainly, because the case for grounding is usually argued in terms of trust. The interviews suggest why. Participants described the trace and the annotations as making a finding easier to locate and to understand, and two of them separated that from believing the assessment. Trust in the verdict was anchored to their own review of the code rather than to the quality of the explanation. Grounding appears to serve comprehension rather than belief.

\subsection{Threats to Validity}
\label{sec:eval-threats}

We group the threats by the standard categories, beginning with the most consequential.

\subsubsection{Construct Validity}
The most important threat is the nature of the independent variable. The user interface, the interaction structure, the axis along which explanations are tiered, the explanation content, and the generation pipeline all differ between the two conditions at once. Every hypothesis therefore attributes its effect to the SAFE-Red tool as a bundle, set against the SAFE-Blue tool as a bundle, and none of them isolates the agentic backend from the interface redesign. This is an explicit construct-level choice rather than an oversight. An observed difference could be carried by the interface, by the tiered presentation, by the backend, or by any combination of them. Separating those contributions requires a factorial ablation over two hybrid configurations that are not currently engineered and would roughly double the sessions or the sample. We rejected that design on engineering and recruitment grounds.

Two elements of the protocol probe the confound without resolving it. The plugin logs interactions per finding, so an outcome can be checked against what participants opened and how long they spent there. The closing interview asks participants to reason about the counterfactual directly, whether their accuracy or speed would have changed had each tool's content been presented in the other's interface. The answers point to how explanations were organised rather than to how they were produced, citing the fixed sections, the short summary and the reduced clicking (Section~\ref{sec:eval-interviews}). That points towards the interface rather than the backend, and it does not settle the question. The question was asked in 13 of 24 sessions, participants are reasoning about a build they used for fifteen minutes, and self-attribution of cause is weak evidence at the best of times. Neither probe substitutes for the ablation. The telemetry delivers less of this than intended, because the two builds log different events for analogous actions and the panel dwell timer is effectively absent from SAFE-Red.

A second construct-level choice concerns the requirements. The redesign addresses all five at once rather than in order of priority, so the marginal contribution of any one of them cannot be reported. Incremental implementation would have given that, at the cost of a design that does not hold together, since R2 and R3 both depend on the agent's ability to consult the codebase. We chose the coherent artefact over the separable one.

The qualitative coding carries a reliability limitation of its own. Segment-level agreement between the two coders on the four double-coded interviews is $46.9$ per cent, which is low, and it is near zero on every code concerned with presentation because such talk pervades the interviews and the coders marked overlapping spans. The interview themes are therefore reported as an interpretive aid and never as a result on their own. Two smaller construct threats remain. Trust, IDE-workflow integration, and the three explanation-quality sub-constructs were captured with a single Likert item rather than multi-item subscales, so no internal consistency can be computed and each rests on the wording of one question. Triage accuracy further presumes a single ground truth per finding, which the lab package fixes ahead of the sessions.

\subsubsection{Internal Validity}
Carry-over between periods is possible in any crossover design. It is addressed by the counterbalancing of Table~\ref{tab:sequences} and by the plugin-free warm-up rather than eliminated. Participants also worked through more findings in the second block than in the first, so a period effect cannot be ruled out. Assignment to the four sequences was moreover not stratified on security expertise, which leaves the sequences unbalanced on triage skill. Because every hypothesis rests on a within-participant contrast, that imbalance does not bias the tool comparison itself.

\subsubsection{External Validity}
Pre-generating the explanations removed generation latency from the task. This keeps the 15-minute blocks comparable, and it means no participant experienced the wait an agentic backend imposes in normal use. Speed was one of the adoption conditions the workshop raised (T5), so the study cannot speak to it, and the time measure reflects reading and deciding rather than the cost of the tool in practice.

The unpublished application bounds what the contamination control covers. Its source is very unlikely to have been memorised, but the five weakness classes are standard and their idiomatic fixes are familiar to both models. An unpublished codebase removes recall of this code rather than recall of SQL injection. It also costs replicability, since a replication would have to substitute an application of comparable structure.

The sample bounds generalisation further. Twelve of the twenty-four participants reported less than a year of Java experience and fourteen had never used a SAST tool (Section~\ref{sec:eval-results}). That matches the beginner-to-intermediate population the plugin targets, and it limits what the results say about developers who triage findings routinely. Two participants hold management roles outside software development, so a small number of decisions come from outside the intended population. We did not run a sensitivity analysis excluding them.

\subsubsection{Conclusion Validity}
The 15-minute cap per block proved to be the binding constraint on data volume. Only one participant reached all five findings in both blocks, and 124 of the 240 possible decisions were recorded, a median of between two and three per block. The number of decisions therefore varies from one to five across participants and conditions, which makes a per-participant accuracy proportion noisy and gives the finding-level mixed-effects model the leading role in the analysis of triage correctness.

\section{Related Work}\label{sec:related-work}

Three lines of work bear on this paper, covering how analyser output is made usable, how language models have been applied to vulnerabilities, and how agents are built. 

\subsection{Improving Usability of Analyser Output}
Work on the explainability of static analysis predates large language models. Rule graphs make the analyser's reasoning visible by encoding the rules and data flows behind a finding~\cite{9106860}. Reading one presupposes a familiarity with taint analysis that the developers in our studies reported lacking. Surveys of the usability criteria these tools meet find most falling short on exposing internal reasoning, explaining consequences, and connecting a warning to the code it concerns~\cite{10.1145/3533767.3534374, nachtigall2019explaining}. A review of adoption barriers separates obstacles inherent to the technique from shortcomings that tool builders can remove~\cite{10.1145/3691621.3694947}, a split our workshop reproduced and the reason R1--R5 reach beyond the wording of an individual warning. Closest to our work is SECGen~\cite{secgen}, which restructures raw analyser output into interpretable reports organised around what a vulnerability is, where it manifests, why it matters, and how to address it. We adopt that decomposition for our explanation schema (Section~\ref{sec:safe-schema}) and differ from it in three ways. SECGen composes a report from what the analyser reported, whereas our agent consults the repository and the rule definition before answering, so an explanation can name the developer's own methods and call sites rather than describing the weakness class. Their report is a standalone artefact, whereas ours is delivered in the IDE beside the flagged code, tiered by demand and linked step by step to the reported data flow (R1, R2, R4). Finally, SECGen validates the generated text automatically, whereas we evaluate the result with developers who have to act on it.

\subsection{Language Models applied to Vulnerabilities}
One group of approaches detects independently, often positioned as a replacement for static analysis. LLMVulExp combines prompting with fine-tuning to detect vulnerabilities and describe them~\cite{llmvulexp}, and DLAP augments a deep-learning detector with model prompting~\cite{dlap_llm}. Systematic comparisons report that models and analysers make different errors rather than one dominating the other~\cite{gnieciak2025large}. A second group reduces the warnings a developer sees rather than explaining them, extending a long line of classifiers trained to separate actionable warnings from noise~\cite{10.1145/3696352}. Models are now used to filter an analyser's output, as in LLM4SA~\cite{auto_inspect_warnings_llm_how_far_are_we}, or to guide the analysis itself~\cite{li2025irisllmassistedstaticanalysis}. A third group repairs what the analyser reports, as GitHub's Copilot Autofix does for CodeQL alerts~\cite{copilot-autofix} and Snyk's DeepCode AI Fix does for Snyk Code~\cite{deepcodeaifix}. We share with LLM4SA the division of labour that leaves detection with the auditable analyser, and differ from all three groups in what the model is asked to do. It neither judges whether a finding is real nor proposes a patch, because a developer has to decide whether to act before either helps. It consults the repository and the rule that fired rather than a supplied snippet, which is what allows an explanation to name the developer's own methods and call sites (R5). And it consumes SARIF, so it applies to whichever analyser a team already runs.

\subsection{Agentic Systems for Software Engineering.}
Giving a model tools and letting it choose which to call turns a single completion into a multi-step investigation~\cite{yao2023reactsynergizingreasoningacting}. In software engineering the pattern has been applied to writing and repairing code. SWE-agent provides an interface for navigating and editing a repository~\cite{3737916.3739517}, CodeAct makes executable code the action space~\cite{3692070.3694124}, and Aider applies the idea in a developer's terminal~\cite{aider}, all judged by whether the resulting patch passes a test suite. We reuse the mechanism for a different end. Our agent produces an explanation rather than a diff, stays bound to the verdict the rule already issued, and navigates only the artefacts that substantiate the reported data flow.

\section{Conclusion \& Future Work}\label{sec:conclusion}

Static analysis is worth defending because its verdicts can be checked. Handing detection to a language model gives that up, and it leaves a developer with nothing outside the model against which to test what the model says. We therefore kept detection with the analyser and gave the model only the explaining.

Two formative studies, a preliminary user study with 20 participants and a workshop with 19 practitioners, produced five requirements for what such a tool has to do. The redesigned SAFE plugin follows them. It tiers explanations by demand rather than by an audience label chosen in advance, and ties each step of an explanation to the data flow the analyser reported. It also replaces a single prompt with an agent that consults the repository and the rule that fired before it answers. The agent is constrained so that it can neither modify code nor form a verdict, and it consumes SARIF, so it applies to whichever analyser a team already runs.

A crossover experiment with 24 developers compared the redesign against its predecessor. Perceived usability rose from $77.3$ to $85.0$ on the System Usability Scale. Actionability showed the widest gain among the explanation items, the aspect both formative studies had identified as weakest. Participants proposed better remediations, with the odds of a correct fix $2.6$ times higher. Triage accuracy and time per finding were indistinguishable. We read those nulls as uninformative rather than as equivalence, because a fifteen-minute block left most participants with two or three decisions. The comparison is between two whole designs, so it establishes that the redesign performs differently rather than isolating which of its parts is responsible.

Two results point beyond this design. Trust in the assessment did not move while every other perception measure did. The interviews indicate why, since participants described the grounding artefacts as making a finding easier to locate and understand rather than easier to believe. Grounding appears to serve comprehension rather than belief, which is not how the case for it is usually argued. Second, participants raised a boundary that no requirement anticipated. It came up in 22 of 24
interviews, and 20 of them drew it in the same place. The tool should find and explain, and the developer should judge and apply. They defended it by pointing to false positives, to the understanding that reviewing a finding gives them, and to being liable for their own code.

\subsection{Future Work}

Four directions follow as future work:

\begin{itemize}
    \item \emph{An ablation that separates the bundle.} Crossing each interface with each backend is the only design that will settle which part of the redesign the improvements came from, and this study cannot substitute for it.
    \item \emph{A task budget that lets accuracy be measured.} Fifteen minutes left most participants with two or three decisions, so the accuracy result is uninformative rather than negative. A longer session or a smaller finding set would give a per-participant proportion worth testing.
    \item \emph{Explanations designed for comprehension rather than persuasion.} If grounding serves understanding rather than belief, an explanation should help a developer reach their own verdict rather than make the analyser's more convincing. What that means concretely is open.
    \item \emph{Support for the boundary developers drew.} They want to judge and apply while the tool finds and explains. Whether an explanation can be built to support that judgement, rather than to substitute for it, is the problem we would take up next.
\end{itemize}

\backmatter

\section*{Declarations}

\bmhead{Funding} This work was supported by the state of North Rhine-Westphalia through its Ministry of Economic Affairs, Industry, Climate Action and Energy (MWIKE) under the project Cyberresilience.nrw.

\bmhead{Competing interests} The authors declare no competing interests.

\bmhead{Acknowledgements} We thank Judy Kayali for her assistance in preparing the materials and processing the results for the industry feedback formative study and Linus Bennin for supporting the development of the agentic backend for the plugin.

\bmhead{Ethics approval and consent to participate}
All participants gave written and verbal consent
before taking part, covering the collection of questionnaire responses, plugin telemetry,
and session recordings, and separately covering the use of pseudonymised quotations in
publications. Participation was voluntary and unpaid, and participants could pause or
withdraw at any point.

\bmhead{Consent for publication}
Participants consented separately to the use of pseudonymised quotations from their
think-aloud and interview transcripts in scientific publications. Only participants who gave
that consent are quoted.

\bmhead{Data availability} The de-identified quantitative data underlying every result in
Section~\ref{sec:eval-results} are available in the replication package \cite{safe}. This covers the pre-study questionnaire responses, the per-finding verdicts and the
ground truth they are scored against, the post-block usability and perception ratings, the
decision durations recovered from the session transcripts, the interaction telemetry, and
the coded fix families together with the resolved codes and the agreement statistics.

The interview and think-aloud transcripts and the session screen recordings are not
available. Participants consented to the publication of aggregated results, de-identified
summary data, and pseudonymised quotations, and not to the release of the recordings or the
raw transcripts. Those materials are held on Fraunhofer IEM encrypted institutional storage, are
accessible only to the named investigators, and are retained for five years after
publication. The qualitative codebooks and the derived counts of codes per participant are
included in the replication package, so the coding can be inspected even though the source
text cannot be released.

\bmhead{Materials availability}
The study instruments are reproduced in full in Appendices~\ref{secA1} and~\ref{secA1b}, and
the fix-family coding scheme in Appendix~\ref{secA2}. The Java application used in the tasks
is internal training material at Fraunhofer IEM and cannot be released, which is a deliberate
choice discussed in Section~\ref{sec:eval-threats} and a cost to replicability. The rule
identifiers and CWE classes of the ten findings are listed in the replication package, so an
equivalent task set can be reconstructed on a substitute application of comparable structure.

\bmhead{Code availability} The SAFE plugin and the agent backend are open source under the MIT licence at
\url{https://github.com/fraunhofer-iem/safe}. The scripts that produce every table and figure in
Section~\ref{sec:eval-results}, including the transcript parsing, the coding-sheet
generation, the agreement statistics, and the statistical tests, are included in the
replication package.

\bmhead{Author contributions}
O.J.\ wrote the main manuscript text and prepared the figures. O.J.\ and R.K.\ designed the methodology, developed the tooling, and conducted the study sessions. O.J., R.K.\ and R.K.S.\ prepared the study materials, recruited the participants, and analysed the data. O.J. and R.K.\ coded the qualitative data. R.K., R.K.S., G.P.\ and E.B.\ reviewed and edited the manuscript. All authors read and approved the final manuscript.

\begin{appendices}

\section{Study Instruments}\label{secA1}

This appendix gives the instruments as they were administered, so that every number in
Section~\ref{sec:eval-results} can be traced back to the question that produced it. Each
item is mapped to the dependent variable it serves.

\subsection{Pre-Study Questionnaire}\label{secA1-pre}

Administered once, before the warm-up. Every item was a single choice from the options
shown.

\begin{table}[h]
\caption{Pre-study questionnaire. The options are those offered, with the ones that no
participant selected marked as unused.}
\label{tab:app-pre}
\begin{tabular}{@{}p{0.46\linewidth}p{0.48\linewidth}@{}}
\toprule
Item & Response Options \\
\midrule
What is your current primary job role? &
Software Developer, Student, Research, Security Engineer/Analyst, Other \\
How many years of professional software development experience do you have? &
$<$1 year, 1--3 years, 3--7 years, 8+ years \\
How many years of experience do you have programming in Java? &
$<$1 year, 1--3 years, 3--7 years (unused), 8+ years \\
How frequently do you use JetBrains products (for example IntelliJ IDEA, PyCharm,
WebStorm, PhpStorm)? &
Never, Rarely, Weekly, Daily \\
How would you rate your familiarity with secure coding practices (for example input
validation, authentication, data protection, database security)? &
Five-point scale, 1 Novice to 5 Expert (5 unused) \\
How frequently do you use Static Application Security Testing (SAST) tools (for example
Semgrep, SonarQube, Qodana)? &
Never, Occasionally, Weekly, Daily \\
Have you ever remediated software vulnerabilities (for example SQL injection,
cross-site scripting, path traversal)? &
No, Yes occasionally, Yes frequently \\
\bottomrule
\end{tabular}
\end{table}

\subsection{Per-Finding Response Form}\label{secA1-form}

Completed once per finding during each task block, for as many of the five findings as the
participant reached.

\begin{table}[h]
\caption{Per-finding response form.}
\label{tab:app-form}
\begin{tabular}{@{}p{0.30\linewidth}p{0.40\linewidth}p{0.22\linewidth}@{}}
\toprule
Item & Response & Serves \\
\midrule
Which finding is this? & The rule identifier shown in the plugin & Item identity \\
What is your final verdict on this finding? & True positive, false positive, or unsure & H1a \\
How would you fix or improve the code? & Free text & H1b \\
\bottomrule
\end{tabular}
\end{table}

The fix item was free text rather than a choice from the fix-family vocabulary, which is
why the proposals had to be coded afterwards. Section~\ref{secA2} gives the coding scheme.

\subsection{Post-Block Questionnaire}\label{secA1-post}

Administered immediately after each of the two task blocks, so that each participant
completed it twice, once per tool.

The System Usability Scale~\cite{sus} was used unmodified, with the word \emph{tool} in
place of \emph{system}. All ten items take a five-point agreement scale from strongly
disagree to strongly agree, and are scored in the standard way to a single value between 0
and 100.

\begin{table}[h]
\caption{System Usability Scale items, in the order presented. Odd items are positively
worded and even items negatively worded, as the instrument requires.}
\label{tab:app-sus}
\begin{tabular}{@{}rp{0.86\linewidth}@{}}
\toprule
\# & Item \\
\midrule
1 & I think that I would like to use this tool frequently. \\
2 & I found the tool unnecessarily complex. \\
3 & I thought the tool was easy to use. \\
4 & I think that I would need the support of a technical person to be able to use this tool. \\
5 & I found the various functions in this tool were well integrated. \\
6 & I thought there was too much inconsistency in this tool. \\
7 & I would imagine that most people would learn to use this tool very quickly. \\
8 & I found the tool very cumbersome to use. \\
9 & I felt very confident using the tool. \\
10 & I needed to learn a lot of things before I could get going with this tool. \\
\bottomrule
\end{tabular}
\end{table}

The five perception items follow the SUS block. Each takes the same five-point
agreement scale. The requirement column names the workshop requirement the item was written
to probe, which is how each one earns its place rather than aggregating into a single
satisfaction score.

\begin{table}[h]
\caption{Perception items, with the hypothesis and requirement each serves.}
\label{tab:app-perception}
\begin{tabular}{@{}p{0.60\linewidth}p{0.13\linewidth}p{0.13\linewidth}@{}}
\toprule
Item & Hypothesis & Requirement \\
\midrule
\emph{Clarity:} The vulnerability explanations provided by the tool were easy to
understand. & H3a & R1 \\
\emph{Actionability:} The tool provided clear next steps to fix the vulnerabilities. &
H3a & R3 \\
\emph{Confidence-support:} The explanations made me feel confident in my decision to
classify the finding as a true or false positive. & H3a & R5 \\
The user interface (tool window, gutter icons, and so on) helped me locate the issues
efficiently. & H3c & R4 \\
I trust the tool's assessment of the vulnerabilities. & H3d & R5 \\
\bottomrule
\end{tabular}
\end{table}

A final item asked, on a scale of 0 to 10, how likely the participant would be to recommend
the plugin to a colleague. It is reported descriptively and carries no hypothesis.

\section{Interview Guide}\label{secA1b}

The closing interview was semi-structured. The five questions below were scripted, and
interviewers followed up freely. Because the script was not applied uniformly, the last column gives the number of the 24
sessions in which each question was actually asked. That is the denominator for any claim
resting on it.

\begin{table}[h]
\caption{Closing interview guide, with the construct each question probes and the number of
sessions in which it was asked.}
\label{tab:app-interview}
\begin{tabular}{@{}rp{0.56\linewidth}p{0.16\linewidth}r@{}}
\toprule
\# & Question & Probes & Asked \\
\midrule
1 & What did you like, not like, or wish to be improved in the usability of the plugins? &
Usability, open & 20 \\
2 & If the information from the SAFE-Red plugin were presented in the SAFE-Blue
audience-toggle interface, do you think your accuracy or speed would have changed, and vice
versa? Why? & Tool-bundle confound & 13 \\
3 & In what way did the evidence links or the code annotations contribute more to your
trust in the tool? & Trust, R5 & 11 \\
4 & Do you think it is necessary to explain vulnerabilities to developers? In the age of
AI, should problems be triaged automatically, and how involved should developers be? &
Automation boundary & 17 \\
5 & Any additional feedback, suggestions, or things you would like to change. & Open & 18 \\
\bottomrule
\end{tabular}
\end{table}

\section{Fix-Family Coding Scheme}\label{secA2}

H1b asks whether the redesign changes the quality of the remediations participants propose. The measure records how well each proposal matches the fix the weakness actually requires. The response form took free text, and roughly half of those fields were left empty. Each
proposal was therefore taken from the written response where one exists, and from the
participant's spoken proposal in the session transcript otherwise.
Coding covers the 64 findings that the ground truth marks as true positives, since a fix
family is only defined where there is something to fix.

\begin{table}[h]
\caption{The expected fix family for each weakness in the material.}
\label{tab:app-families}
\begin{tabular}{@{}llp{0.52\linewidth}@{}}
\toprule
Family & Weakness & Expected mechanism \\
\midrule
F1 & Path traversal (CWE-23) & Canonicalise the resolved path and check that it stays
inside the allowed base directory \\
F2 & SQL injection (CWE-89) & Prepared statement, parameterised query, or ORM binding \\
F3 & Server-side request forgery (CWE-918) & Constrain the host through an allowlist, or a
trusted base URL with only the path and query under user control \\
F4 & Cross-site scripting (CWE-79) & Render through a view or template with contextual
output encoding, or escape before writing \\
\bottomrule
\end{tabular}
\end{table}

\begin{table}[h]
\caption{The four levels of the fix-family measure and the rules that separate them.}
\label{tab:app-levels}
\begin{tabular}{@{}lp{0.34\linewidth}p{0.44\linewidth}@{}}
\toprule
Level & Meaning & Decision rules \\
\midrule
\texttt{correct} & Names the right mechanism for the family & Includes restating the fix
the tool suggested, and naming the safe counterpart method \\
\texttt{partial} & Right family, mechanism vague or incomplete & Sanitisation or validation
language with no named mechanism \\
\texttt{incorrect} & Wrong family, or a mechanism that does not address the weakness &
A fix belonging to a different family \\
\texttt{none} & No remediation proposed & Commentary about the tool rather than the code,
proposing detection rather than a fix, and declining to fix because the finding was judged
not real \\
\bottomrule
\end{tabular}
\end{table}

Two coders applied the scheme blind to participant and condition. They calibrated on 15
items and revised the codebook, then coded the remainder with 20 items in common.
Quadratic weighting was used for the agreement statistic because the levels are ordinal, so
that a disagreement between \texttt{correct} and \texttt{partial} counts for less than one
between \texttt{correct} and \texttt{none}. Agreement on the 20 post-revision items was 80
per cent exact with a quadratic-weighted Cohen's kappa of $0.79$. The eight disagreements
across all double-coded items were settled by discussion.

The three rules in the last column of Table~\ref{tab:app-levels} that concern
\texttt{none} were added after calibration. Each resolved a case the original scheme left
undefined, and together they accounted for most of the disagreement in the calibration
round.

\section{Interview Coding Scheme}\label{secA3}

The closing interviews were coded with the hybrid scheme below. The five requirements
served as sensitising concepts, and open codes covered what they did not anticipate. Two
coders built the code system together on five transcripts, then applied it independently,
with four transcripts coded by both. Table~\ref{tab:app-jobb} lists the resulting codes
with the number of participants and segments each covers, over 211 segments and all 24
interviews.

\begin{table}[h]
\caption{Interview codes, ordered by how many participants support each. The last column
gives segment-level agreement on the four double-coded interviews, where a dash marks a code
with too few compared codings to report.}
\label{tab:app-jobb}
\begin{tabular}{@{}p{0.30\linewidth}rrp{0.30\linewidth}@{}}
\toprule
Code & Part. & Seg. & Agreement \\
\midrule
Automation boundary & 22 & 40 & 67 per cent \\
Explanation structure and content & 21 & 38 & 0 per cent \\
Expertise level & 17 & 23 & 67 per cent \\
Importance of explaining & 14 & 19 & 67 per cent \\
Fix suggestions & 13 & 17 & 50 per cent \\
Data-flow navigation & 12 & 19 & 67 per cent \\
Code annotations & 12 & 15 & 67 per cent \\
Plugin interface & 7 & 10 & 0 per cent \\
Explanation clarity & 6 & 8 & 0 per cent \\
Length of text & 6 & 7 & 0 per cent \\
Development workflow & 4 & 6 & 0 per cent \\
TL;DR & 3 & 4 & --- \\
Deep dive & 3 & 3 & --- \\
Code examples & 2 & 2 & --- \\
\midrule
Overall & 24 & 211 & 46.9 per cent \\
\bottomrule
\end{tabular}
\end{table}

Segments were selected as spans carrying a complete claim and were allowed to cross turn
boundaries, since the transcription tool splits mid-sentence. One segment can carry several
codes. Interviewer questions, procedural talk and bare acknowledgements were not coded, and
neither was any passage whose speaker could not be determined. Agreement was computed at
segment level in MAXQDA \cite{radiker2019analyse} on the four double-coded interviews, over 64 compared codings. The
figure MAXQDA reports as kappa uses the Rädiker and Kuckartz definition of chance agreement,
which is not meaningful for a code system of this size, so percentage agreement is reported
instead.

The pattern in the last column is itself informative. Codes with a concrete referent in
the interface or the task agree at 67 per cent, while codes about presentation agree at zero.
Talk about structure, clarity, length and the interface runs through every answer rather than
sitting in identifiable passages, so the two coders marked overlapping but non-coincident
spans, which the overlap criterion counts as disagreement. Section~\ref{sec:eval-interviews}
reports the themes accordingly, as an aid to interpreting the quantitative results rather
than as findings in their own right.

\end{appendices}


\bibliography{sn-bibliography}

\end{document}